\documentclass[a4paper,11pt]{article}
\usepackage{pos}
\usepackage{graphicx}
\usepackage{subcaption}
\usepackage{hyperref}
\usepackage{caption}
\usepackage{float}
\usepackage{comment}
\definecolor{myblue}{RGB}{0,0,128}
\usepackage{makecell}

\title{Application of the optimised next neighbour image cleaning method to the VERITAS array}
\ShortTitle{Application of the optimised next neighbour image cleaning method for the VTS array}

\author{Maria Kherlakian$^{a, b, *}$, on behalf of the VERITAS Collaboration}

\affiliation[a]{Deutsches Elektronen-Synchrotron DESY,\\
  Platanenallee 6, 15738 Zeuthen, Germany}

\affiliation[b]{Humboldt-Universität zu Berlin, Faculty of Mathematics and Natural Sciences,\\
Newtonstraße 15, 12489 Berlin, Germany}

\emailAdd{maria.kherlakian@desy.de}

\abstract{Imaging atmospheric Cherenkov telescopes, such as the VERITAS array, are subject to the Night Sky Background (NSB) and electronic noise, which contribute to the total signal of pixels in the telescope camera. The contribution of noise photons in event images is reduced with the application of image cleaning methods. Conventionally, high thresholds must be employed to ensure the removal of pixels containing noise signal. On that account, low-energy gamma-ray showers might be suppressed during the cleaning. We present here the application of an optimised next neighbour image cleaning for the VERITAS array. With this technique, differential noise rates are estimated for each individual observation and thus changes in the NSB and afterpulsing are consistently being accounted for. We show that this method increases the overall rate of reconstructed gamma-rays, lowers the energy threshold of the array and allows the reconstruction of low energy (E $>$ 70 GeV) source events which were suppressed by the conventional cleaning method.}

\ConferenceLogo{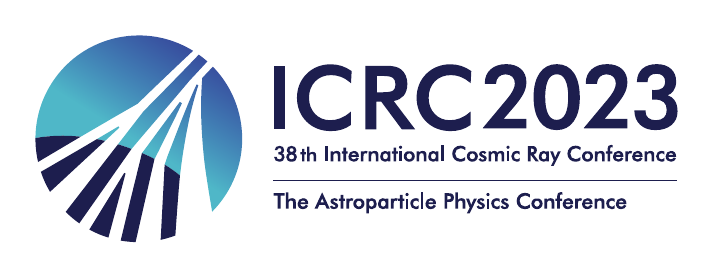}

\FullConference{%
38th International Cosmic Ray Conference (ICRC2023)\\
  26 July - 3 August, 2023\\
  Nagoya, Japan}

\begin{document}
\maketitle

\section{Introduction}

Imaging atmospheric Cherenkov telescopes (IACTs) are instruments designed to capture the faint Cherenkov light emitted by high-energy particles produced by the interaction of gamma-ray photons with the Earth atmosphere. The Very Energetic Radiation Imaging Telescope Array System (VERITAS) is an array of four IACTs in operation since 2007 and located at the Fred Lawrence Whipple Observatory (FLWO) in southern Arizona (31 40N, 110 57W,  1.3km a.s.l.). With a field of view of 3.5$^{\circ}$, it can detect very-high-energy (VHE) gamma-ray sources with the strength of 1$\%$ Crab Units (C.U.) in $\sim$ 25 hours \cite{veritas}. 

IACTs collect not only the Cherenkov light from particle showers, but also an undesired amount of noise photons originating from the NSB with an additional interference introduced by the electronics and detector. When telescopes are triggered, the recorded image contains not only the true Cherenkov signal but also noise pixels, that, if not removed, introduce a strong bias in image parametrization. The noise level in a given data collection is estimated through pedestal events, which are artificial triggers of the camera readout performed at a rate of 1 Hz throughout an observation. In this way, the signal in the camera pixels is known in the absence of Cherenkov light. Traditionally, noise is removed from telescope images by applying threshold cuts to the integrated signal collected per pixel. The final image to undergo parametrization is composed of pixels containing an integrated signal that is at least 5 times the standard deviation of their pedestal (\textit{pedvar}) and whose neighbouring pixels are at least 2.5$\sigma$ above their own pedestal. These thresholds are always applied regardless of the noise sources of a given observation, thus ensuring that mostly pixels with Cherenkov signal are retained for the event reconstruction. Gradients in the signal arrival times are taken into account by the application of the double pass method \cite{veritas}. Although this technique of image cleaning is successful in removing noise-type pixels, images produced by low energy showers can be suppressed by such high signal cutoffs, resulting in an increase of the energy threshold of the telescope. 

One way to ensure that low-charge pixels are not bypassed for the event reconstruction is to explore not only the shower signal, but also the time structure of a Cherenkov pulse in the camera. An example of a cleaning method that makes use of this concept is the optimised next neighbour~\cite{NN} technique, which focuses on suppressing bright pixels originating from NSB photons or electronic noise. With this strategy, changes in the noise are consistently being accounted for, which ensures that optimal cutoffs are applied to each observation. Although this method has been successfully tested in Monte Carlo simulations for the Cherenkov Telescope Array (CTA) experiment and widely used, for example, for the estimation of CTAO sensitivities, this is the first time that its performance is studied on IACT data. A simplified approach in which time information is also taken into account has been introduced by the MAGIC array of telescopes and showed remarkable gains in terms of event reconstruction~\cite{magic}.

\section{Method description: the Optimised Next Neighbour cleaning}

\begin{figure}[ht]
   \begin{minipage}[t]{0.48\textwidth}
     \centering
     \includegraphics[width=.9\linewidth]{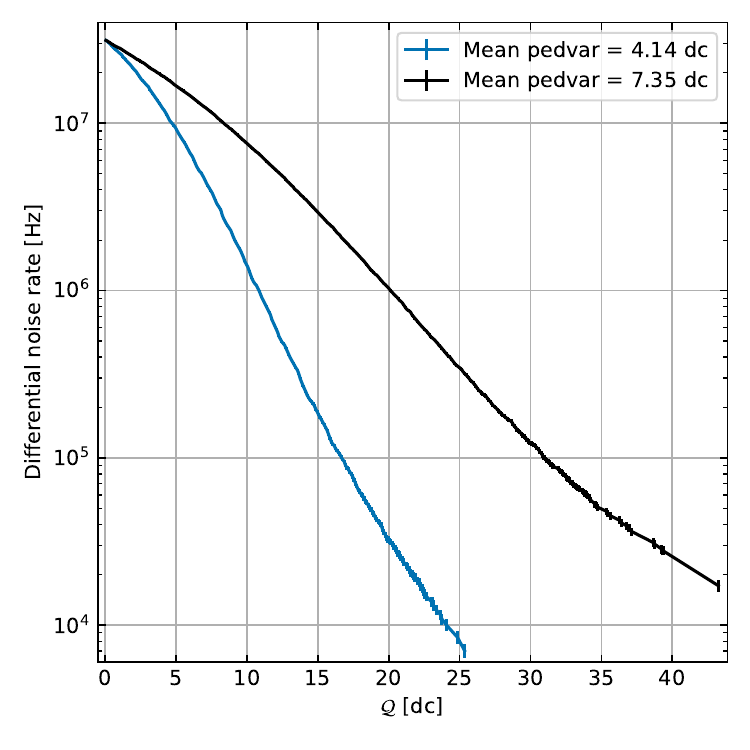}
     \caption{Differential noise rate per pixel signal $\mathcal{Q}$ in digital counts (dc) for observations with \textit{pedvar} = 7.35 dc and \textit{pedvar} = 4.14 dc. Curves are calculated for a camera read-out window time of 32.0 ns and integration window of 12.0 ns.}\label{fig:ipr}
   \end{minipage}\hfill
   \begin{minipage}[t]{0.48\textwidth}
     \centering
     \includegraphics[width=.9\linewidth]{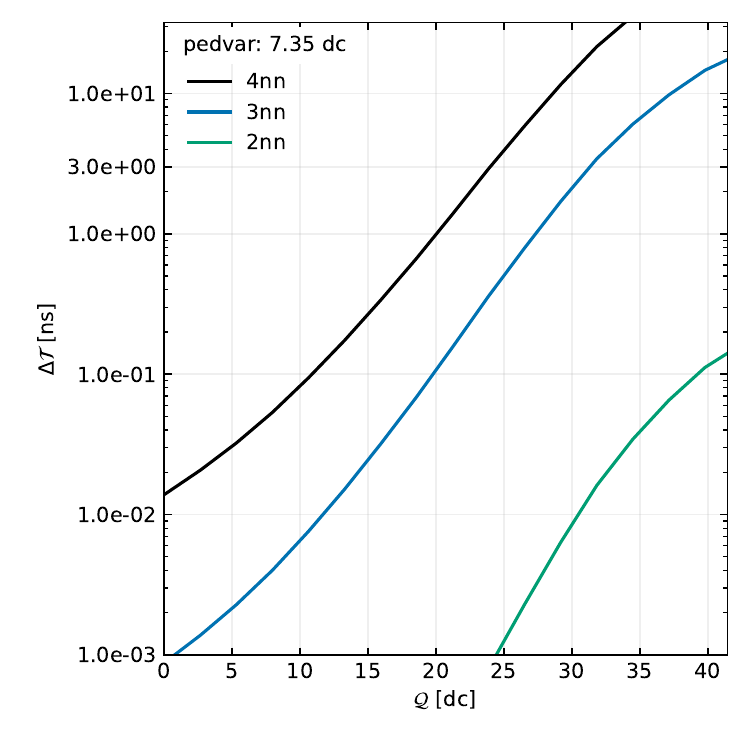}
     \caption{Contours of maximum time coincidence window, $\Delta\mathcal{T}$, and minimum group charge, $\mathcal{Q}$, for 4nn, 3nn and 2nn multiplicities. The plot is limited to the size of the read-out window (32 ns).}\label{fig:contours}
   \end{minipage}
\end{figure}

In the Optmized Next Neighbour method, groups of neighbouring pixels of different multiplicities (2 next-neighbours (2nn), 3nn or 4nn) must meet a maximum time coincidence window, $\Delta\mathcal{T}$, for a given minimum group pixel charge, $\mathcal{Q}$, so that there is a maximum probability that the group is of noise type rather than originating from the Cherenkov light of the shower. In this way, the cleaning cuts are calculated dynamically, are independent of the gamma-ray sources, and can be computed through pedestal events only. The maximum accidental rate, $\mathcal{R}_{acc}$, at which shower images will present neighbouring pixels with noise signal is given by:
%\vspace{-0.01cm}
\begin{equation}
{R}_{acc}(\mathcal{Q}, \Delta \mathcal{T}, n) = C_{n} \Delta \mathcal{T} ^{n-1} \mathcal{I}_{pix}( \mathcal{Q} )^{n},
\label{eq:contours}
\vspace{-0.2cm}
\end{equation}
where $C_{n}$ is the number of permutations of multiplicity $n$ given the geometric distribution of pixels in the camera and $\mathcal{I}_{pix}$ is the typical differential noise rate at which pixels have a signal above a given charge threshold, $\mathcal{Q}$. Assuming that all pixels are identical, one can average $\mathcal{I}_{pix}$ across all channels of each telescope camera. An example of $\mathcal{I}_{pix}$ for the VERITAS array obtained under two different NSB conditions is shown in Figure \ref{fig:ipr}. In this particular example, the curve was obtained in reduced high-voltage mode for \textit{pedvar} = 4.14 dc and in dark conditions for \textit{pedvar} = 7.35 dc. By inverting Equation \ref{eq:contours}, it is possible to derive the dynamic cuts in the parameter space $\Delta\mathcal{T}$ - $\mathcal{Q}$. Examples of contours calculated for a typical 20 minutes run and assuming a probability of 0.05$\%$, i.e., ${R}_{acc}$ = 3125.0 Hz, for noise generated groups of pixels, are shown in Figure \ref{fig:contours}.

\section{Results}
 
\begin{figure}
        \centering
        \begin{subfigure}[t]{0.475\textwidth}
            \centering
            \includegraphics[width=0.8\textwidth]{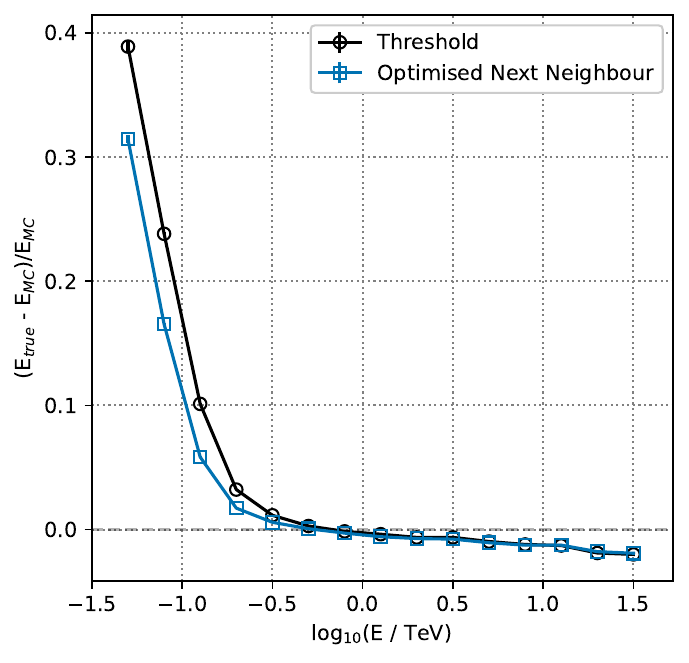}
            \caption{The energy bias, ($E_{reco}$ - $E_{MC}$)/$E_{MC}$, where $E_{reco}$ is the reconstructed energy and $E_{MC}$ the energy of the simulated event.}   
            \label{fig:engbias}
        \end{subfigure}
        \hfill
        \begin{subfigure}[t]{0.475\textwidth}  
            \centering 
            \includegraphics[width=0.8\textwidth]{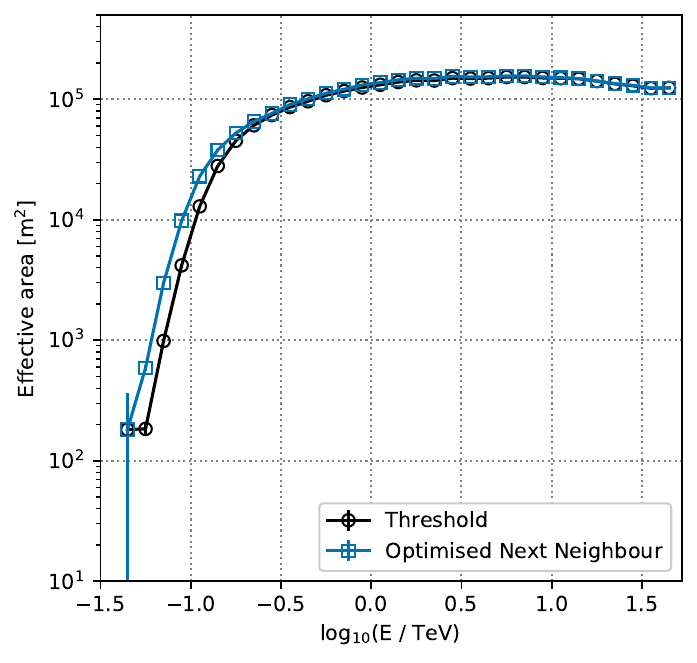}
            \caption{Effective areas as a function of energy.}   
            \label{fig:effarea}
        \end{subfigure}
        \vskip\baselineskip
        \begin{subfigure}[t]{0.475\textwidth}   
            \centering 
            \includegraphics[width=0.8\textwidth]{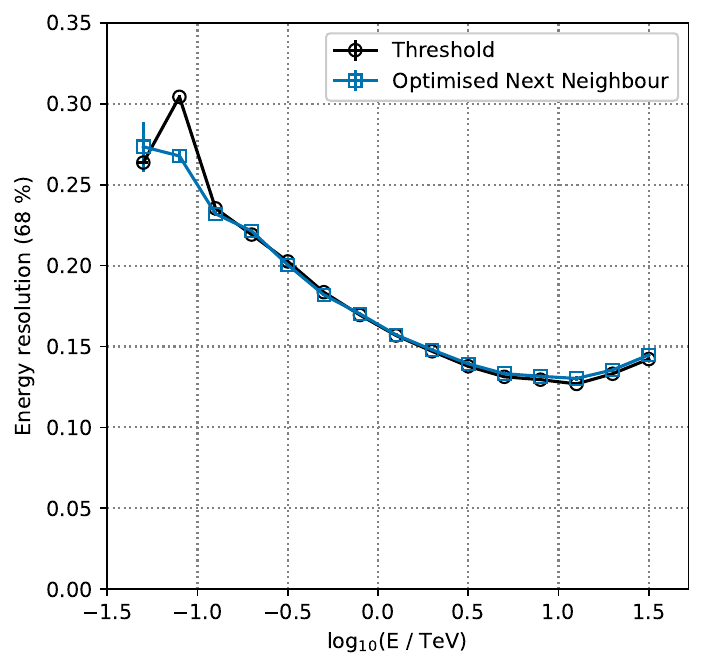}
            \caption{Energy resolution (68$\%$ confinement level).}
            \label{fig:engres}
        \end{subfigure}
        \hfill
        \begin{subfigure}[t]{0.475\textwidth}   
            \centering 
            \includegraphics[width=0.8\textwidth]{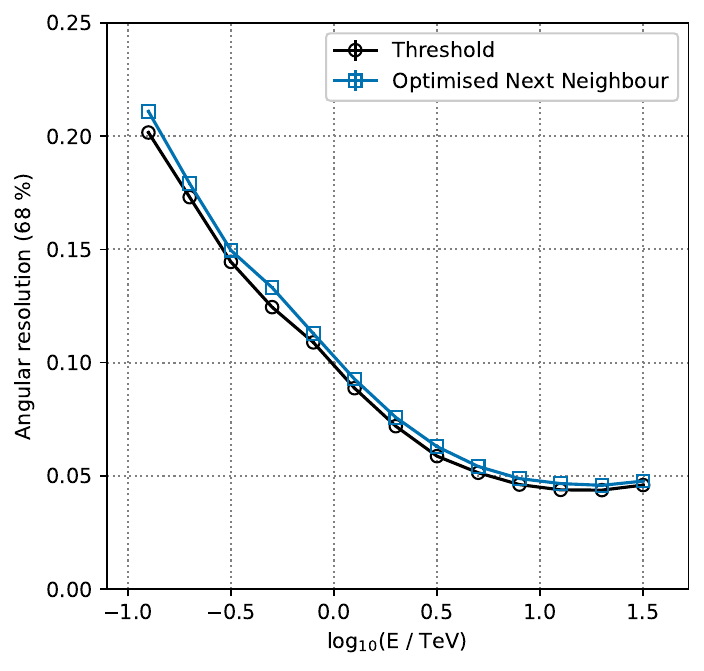}
            \caption{Angular resolution (68$\%$ confinement level).}   
            \label{fig:angres}
        \end{subfigure}
        \caption{Instrument response functions for the optimised next neighbour (blue) and threshold (TS) techniques: energy bias, effective areas, energy and angular resolutions for simulated showers from the north direction, at zenith 30$^{\circ}$ and noise rate 200 MHz. Simulations are produced considering winter conditions for the 2012-2013 observing season.}
        \label{fig:irfs}
\end{figure}

 The performance of optimised next neighbour (NN) cleaning for the VERITAS array was first evaluated using Monte Carlo simulations produced with the CARE\footnote{\href{https://github.com/nepomukotte/CARE}{https://github.com/nepomukotte/CARE}} package. The instrument response functions (IRFs) were compared to those obtained using the threshold technique (TS). Figure \ref{fig:irfs} illustrates the energy bias, \ref{fig:engbias}, effective area, \ref{fig:effarea}, energy, \ref{fig:engres}, and angular resolutions, \ref{fig:angres}, for shower simulations from the north direction at 30$^{\circ}$ zenith, and 200 MHz noise rate. By exploring the time structure of the pulse arrival in the camera, the NN method can avoid the suppression of low signal images that would have been cut off by the standard cleaning. Consequently, a higher number of low energy showers is reconstructed. As a result, the telescope’s energy threshold is reduced, due to an increase in effective areas at lower energies. Because the threshold method performs well in reconstructing showers generated by high energy particles, no changes besides statistical fluctuations are observed in IRFs above $\sim$ 300 GeV. As expected, the energy and angular resolution, obtained as in \cite{performance}, produced similar results for both techniques.

We also evaluate the performance of the method using VERITAS array data. For this purpose, we compare the results obtained from the analysis of three gamma ray sources: the Crab Nebula, which serves as a calibration source due to its well-characterized gamma-ray emisison, and the High Frequency BL-Lacs (HBLs) PKS 1424+240 and PG 1553+113. The investigated HBLs have high spectral indices and are therefore suitable sources for testing the method at low energies. For PKS 1424+240, spectral indices of $\Gamma$ = 3.8 $\pm$ 0.3 and $\Gamma$ = 4.3 $\pm$ 0.3 are obtained respectively for the 2009 and 2011 VERITAS observations, in which the source achieved 4.6$\%$ of the Crab flux above 120 GeV. During the dimmer VHE state in 2013, in which the HBL achieved 2.2$\%$ of the Crab flux above 120 GeV, observations yield $\Gamma$ = 4.5 $\pm$ 0.2 \cite{pks1424_2}. By combining the three epochs, a spectral index of $\Gamma$ = 4.2 $\pm$ 0.3 \cite{pks1424_2} is found. For PG 1553+113, the spectral index of $\Gamma$ = 4.33 $\pm$ 0.09 is obtained via VERITAS observations spanning from 2010 to 2012 \cite{pg1553_2}. For a total exposure time, T$_{\text{exp}}$, of data taken in 0.5$^{\circ}$ wobble mode, the number of ON and OFF events is calculated with background modelled as reflected regions \cite{berge2007}. The gamma/hadron separation is done with cuts optimised for the highest sensitivity for a soft source ($\Gamma$ = 3) and to have the lowest possible energy threshold. Table \ref{tab:analysis} shows the number of signal events (ON-OFF), the significance of the detection \cite{li&ma} and the gamma-ray and background rates per minute for each source.

\subsection{Crab Nebula}

\begin{figure}[t]
   \begin{minipage}[t]{0.48\textwidth}
     \centering
     \includegraphics[width=.9\linewidth]{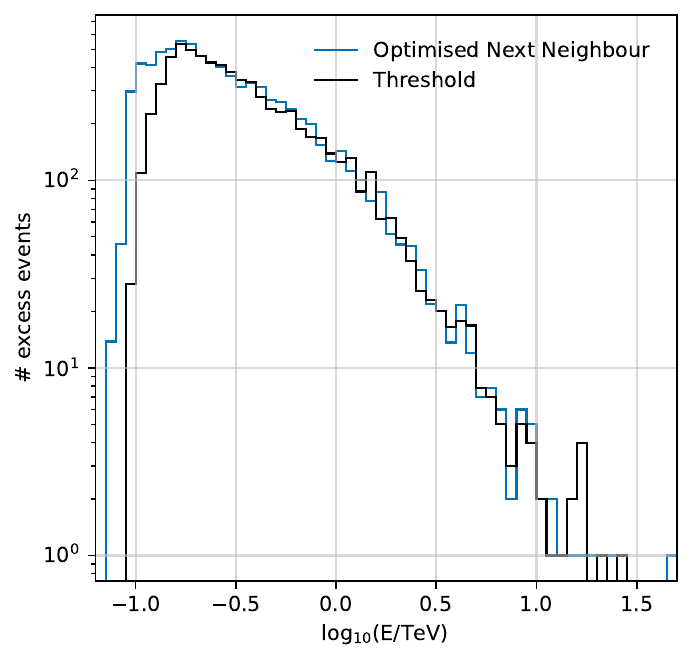}
     \caption{Number of excess events obtained with the optimised next neighbour (blue line) and threshold (black line) cleaning methods for 9.5 hours of Crab Nebula data.}\label{Fig:crab_counts}
   \end{minipage}\hfill
   \begin{minipage}[t]{0.48\textwidth}
     \centering
     \includegraphics[width=.9\linewidth]{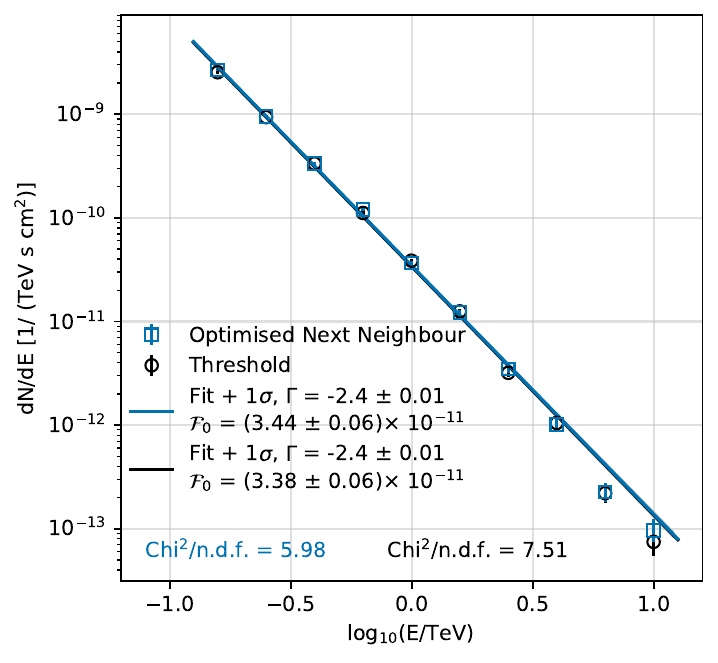}
     \caption{Reconstruction of the Crab Nebula spectrum obtained with the optimised next neighbour (blue line) and threshold (black line) cleaning  methods. Flux constant $\mathcal{N}_{0}$ given in 1/( TeV s cm$^{-2}$) at 1 TeV.}\label{Fig:crab_spectrum}
   \end{minipage}
\end{figure}

 Figure \ref{Fig:crab_counts} presents the number of signal events reconstructed with the NN (blue) and TS (black) cleaning methods in 9.5 hours of Crab Nebula data obtained with the VERITAS array with mean elevation of 69.0$^{\circ}$. The histogram is presented in bins of width 0.05 in logarithmic scale. Below 140 GeV, we observe an increase of at least 48$\%$ per energy bin in the number of signal events reconstructed with the NN method compared to the standard cleaning. Moreover, with the new method, we see reconstructed events in energy bins centred at 80.0 GeV and 90.0 GeV, which are lacking for the TS method. Figure \ref{Fig:crab_spectrum} shows the reconstruction of the energy spectrum in energy bins of width 0.2 in logarithmic scale with at least 10 signal events per spectral point. An increase from 12.2 $\pm$ 0.2 to 14.3 $\pm$ 0.2 gamma-ray events per minute was observed. The significance of detection was reduced by 13 $\sigma$ with NN, which occurs due to the also considerable increase of the background rate. By using gamma/hadron separation cuts optimised with boosted decision trees, a similar detection significance is obtained for the Crab Nebula with both methods and is highly increased for the HBLs analysed here.

\subsection{PKS 1424+240}

\begin{figure}[t]
   \begin{minipage}[t]{0.48\textwidth}
     \centering
     \includegraphics[width=.9\linewidth]{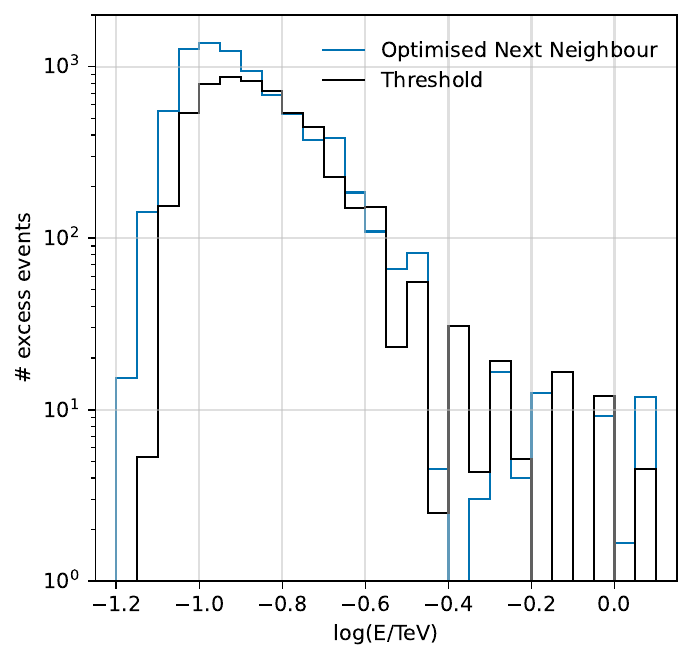}
     \caption{Number of excess events obtained with the optimised next neighbour (blue line) and threshold (black line) cleaning methods for 135.8 hours of exposure time on PKS 1424+240.}\label{Fig:pks1424_counts}
   \end{minipage}\hfill
   \begin{minipage}[t]{0.48\textwidth}
     \centering
     \includegraphics[width=.9\linewidth]{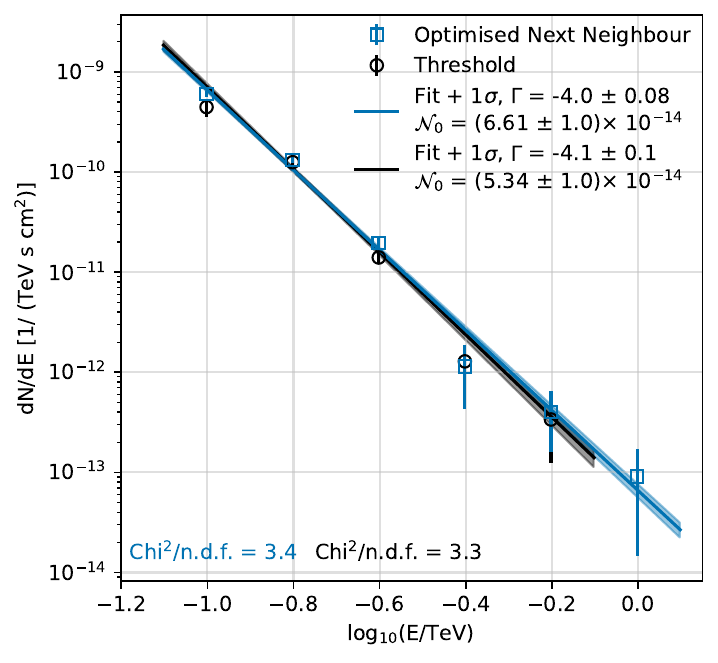}
     \caption{Reconstruction of PKS 1424+240 spectrum  for the optimised next neighbour (blue line) and threshold (black line) cleaning  methods. Flux constant $\mathcal{N}_{0}$ given in 1/(TeV s cm$^{-2}$) at 1 TeV.}\label{Fig:pks1424_spectrum}
   \end{minipage}
\end{figure}

The signal counts spectrum for PKS 1424+240 for T$_{\text{exp}}$ = 135.8 hours and mean elevation of 75.7$^{\circ}$ is shown in Figure \ref{Fig:pks1424_counts}. An overall increase in the number of events is observed, particularly at low energies, with at least 14$\%$ more counts per energy bin below 140 GeV for the NN cleaning. As presented in Figure \ref{Fig:pks1424_spectrum}, the  reconstruction of the source spectrum has one more spectral point at 1 TeV for NN. The spectral index obtained is due to the time period of the analysed dataset, which consists of data taken in the period from 2012 to 2020, in which the source presents different flux states. There is a noticeable increase in the gamma-ray rate, with the number of events per minute rising from 0.66 $\pm$ 0.03 to 0.98 $\pm$ 0.04.

\subsection{PG 1553+113}

\begin{figure}[t]
   \begin{minipage}[t]{0.48\textwidth}
     \centering
     \includegraphics[width=.9\linewidth]{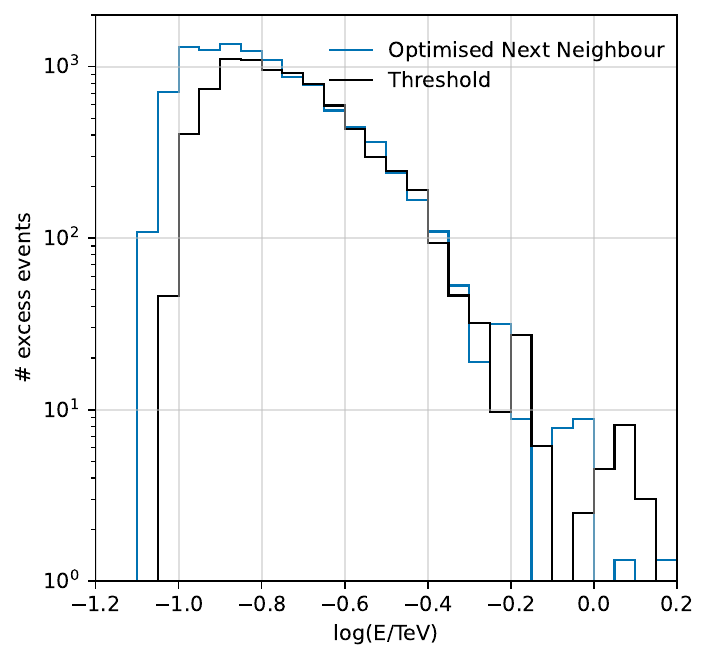}
     \caption{Number of excess events obtained with the optimised next neighbour (blue line) and threshold (black line) cleaning  methods for 63.5 hours of exposure time on PG 1553+113.}\label{Fig:pg1553_counts}
   \end{minipage}\hfill
   \begin{minipage}[t]{0.48\textwidth}
     \centering
     \includegraphics[width=.9\linewidth]{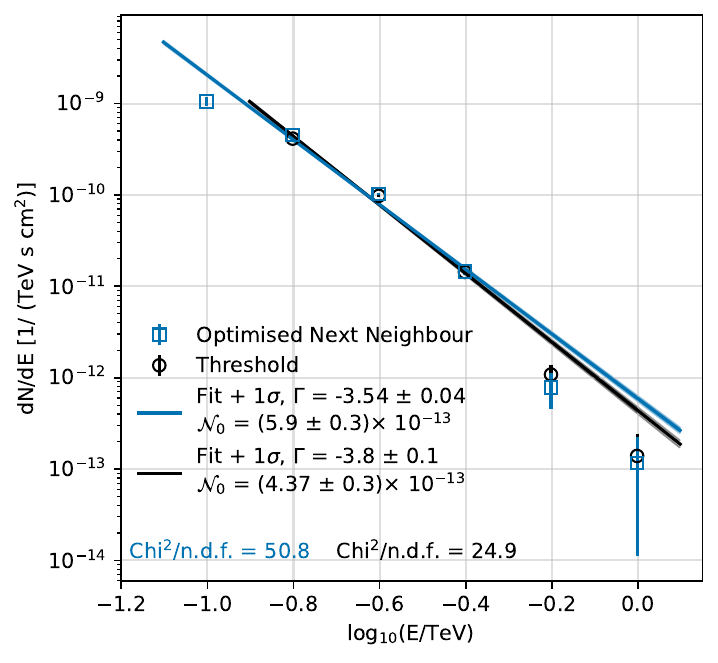}
     \caption{Reconstruction of PG 1553+113 spectrum obtained with the optimised next neighbour (blue line) and threshold (black line) cleaning  methods. Flux constant $\mathcal{N}_{0}$ given in 1/(TeV s cm$^{-2}$).}\label{Fig:pg1553_spectrum}
   \end{minipage}
\end{figure}

For PG 1553+113, with T$_{\text{exp}}$ = 63.5, an increase of at least 23$\%$ per energy bin is observed below 140 GeV with the NN method, as presented in Figure \ref{Fig:pg1553_counts}. Again, we see signal events reconstructed in the 90 GeV energy bin, which is not achieved with the TS method. Figure \ref{Fig:pg1553_spectrum} shows an additional spectral point at 100 GeV. Once more, the spectral indices obtained are explained by the combined dataset, which is composed of observations performed in the period from 2012 to 2018. An increase from 2.11 $\pm$ 0.04 to 2.81 $\pm$ 0.05 events per minute was observed for the gamma-ray rate.

\begin{table}[]
\centering
\begin{tabular}{llcc}
\\ \Xhline{4\arrayrulewidth}
             &                                  & Optimised Next Neighbour & Threshold       \\ \hline
\textbf{Crab Nebula:}       & \multicolumn{1}{l|}{T$_{\text{exp}}$ (min)} &       572.2                  &      572.2        \\
             %& \multicolumn{1}{l|}{ON / OFF}      & 13553 / 5509.2            & 9457 / 3936.7     \\
             & \multicolumn{1}{l|}{ON - OFF}    & 11810.2 & 8526.8 \\
             & \multicolumn{1}{l|}{$S$ $(\sigma)$} & 89.7                    & 102.7            \\
             & \multicolumn{1}{l|}{gamma/min}   & 14.3 $\pm$ 0.2          & 12.2 $\pm$ 0.2  \\
             & \multicolumn{1}{l|}{bkg/min}     & 7.7                    & 3.2            \\ \hline
\textbf{PKS 1424+240:} & \multicolumn{1}{l|}{T$_{\text{exp}}$ (min)} &       8149.82                  &      4860.5   \\
             %& \multicolumn{1}{l|}{ON / OFF}      & 63738 / 59580.83          & 35898 / 32850.83  \\
             & \multicolumn{1}{l|}{ON - OFF}    & 72043.3                 & 31304.9        \\
             & \multicolumn{1}{l|}{$S$ $(\sigma)$} & 26.1                    & 25.6            \\
             & \multicolumn{1}{l|}{gamma/min}   & 0.98 $\pm$ 0.04         & 0.66 $\pm$ 0.03 \\
             & \multicolumn{1}{l|}{bkg/min}     & 9.44                   & 4.75            \\ \hline
\textbf{PG1553+113:}   & \multicolumn{1}{l|}{T$_{\text{exp}}$ (min)} &       3811.3                  &      3811.3   \\
             %& \multicolumn{1}{l|}{ON / OFF}      & 41671 / 31326.00          & 24597 / 16986.17  \\
             & \multicolumn{1}{l|}{ON - OFF}  & 32081.9                  & 18424.5     \\
             & \multicolumn{1}{l|}{$S$ $(\sigma)$} & 57.7                    & 60.0            \\
             & \multicolumn{1}{l|}{gamma/min}   & 2.81 $\pm$ 0.05         & 2.11 $\pm$ 0.04 \\
             & \multicolumn{1}{l|}{bkg/min}     & 6.73                    & 3.27            \\ \Xhline{4\arrayrulewidth}
\end{tabular}
\caption{Summary of analysis results: signal counts (ON - OFF), significance $S$ and gamma-ray and background rates for a total exposure time T$_{\text{exp}}$ with the optimised next neighbour ans threshold cleaning methods.}
\label{tab:analysis}
\end{table}

\section{Summary}

In this study we compare the performance of the optimised next neighbour image cleaning with the standard technique used by the VERITAS array. We show that the former method is able to reduce the energy threshold of the telescope due to an increased number of reconstructed signal events. For all analysed sources, we notice an evident boost in the number of low energy signal events and an overall increase in the gamma-ray rate. The results presented here show that this technique can be of particular relevance for the analysis of soft-spectrum sources, for which the gain in low energies plays an important role.

\section*{Acknowledgements}
This research is supported by grants from the U.S. Department of Energy Office of Science, the U.S. National Science Foundation and the Smithsonian Institution, by NSERC in Canada, and by the Helmholtz Association in Germany. This research used resources provided by the Open Science Grid, which is supported by the National Science Foundation and the U.S. Department of Energy's Office of Science, and resources of the National Energy Research Scientific Computing Center (NERSC), a U.S. Department of Energy Office of Science User Facility operated under Contract No. DE-AC02-05CH11231. We acknowledge the excellent work of the technical support staff at the Fred Lawrence Whipple Observatory and at the collaborating institutions in the construction and operation of the instrument.

\bibliographystyle{unsrt}
\bibliography{references}

%\begin{thebibliography}{99}
%\bibitem{...}
%....
%\end{thebibliography}

%% Full authors list (ONLY FOR COLLABORATIONS)
%\clearpage
%\section*{Full Authors List: \Coll\ Collaboration}
%
%\noindent \textbf{Note comment afterwards:} Collaborations have the possibility to provide an authors list in xml format which will be used while generating the DOI entries making the full authors list searchable in databases like Inspire HEP. For instructions please go to icrc2021.desy.de/proceedings or contact us under icrc2021proc@desy.de.\\
%
%\scriptsize
%\noindent
%first.author$^1$, 
%second.author$^2$, 
%third.author$^3$ % .... more names
%and 
%last.author$^{n}$ \\
%
%\noindent
%$^1$first.affiliation.
%$^2$second.affiliation. % .... more affiliation
%$^{m}$last.affiliation.
\clearpage

\section*{Full Author List: VERITAS Collaboration}

\scriptsize
\noindent
A.~Acharyya$^{1}$,
C.~B.~Adams$^{2}$,
A.~Archer$^{3}$,
P.~Bangale$^{4}$,
J.~T.~Bartkoske$^{5}$,
P.~Batista$^{6}$,
W.~Benbow$^{7}$,
J.~L.~Christiansen$^{8}$,
A.~J.~Chromey$^{7}$,
A.~Duerr$^{5}$,
M.~Errando$^{9}$,
Q.~Feng$^{7}$,
G.~M.~Foote$^{4}$,
L.~Fortson$^{10}$,
A.~Furniss$^{11, 12}$,
W.~Hanlon$^{7}$,
O.~Hervet$^{12}$,
C.~E.~Hinrichs$^{7,13}$,
J.~Hoang$^{12}$,
J.~Holder$^{4}$,
Z.~Hughes$^{9}$,
T.~B.~Humensky$^{14,15}$,
W.~Jin$^{1}$,
M.~N.~Johnson$^{12}$,
M.~Kertzman$^{3}$,
M.~Kherlakian$^{6}$,
D.~Kieda$^{5}$,
T.~K.~Kleiner$^{6}$,
N.~Korzoun$^{4}$,
S.~Kumar$^{14}$,
M.~J.~Lang$^{16}$,
M.~Lundy$^{17}$,
G.~Maier$^{6}$,
C.~E~McGrath$^{18}$,
M.~J.~Millard$^{19}$,
C.~L.~Mooney$^{4}$,
P.~Moriarty$^{16}$,
R.~Mukherjee$^{20}$,
S.~O'Brien$^{17,21}$,
R.~A.~Ong$^{22}$,
N.~Park$^{23}$,
C.~Poggemann$^{8}$,
M.~Pohl$^{24,6}$,
E.~Pueschel$^{6}$,
J.~Quinn$^{18}$,
P.~L.~Rabinowitz$^{9}$,
K.~Ragan$^{17}$,
P.~T.~Reynolds$^{25}$,
D.~Ribeiro$^{10}$,
E.~Roache$^{7}$,
J.~L.~Ryan$^{22}$,
I.~Sadeh$^{6}$,
L.~Saha$^{7}$,
M.~Santander$^{1}$,
G.~H.~Sembroski$^{26}$,
R.~Shang$^{20}$,
M.~Splettstoesser$^{12}$,
A.~K.~Talluri$^{10}$,
J.~V.~Tucci$^{27}$,
V.~V.~Vassiliev$^{22}$,
A.~Weinstein$^{28}$,
D.~A.~Williams$^{12}$,
S.~L.~Wong$^{17}$,
and
J.~Woo$^{29}$\\
\\
\noindent
$^{1}${Department of Physics and Astronomy, University of Alabama, Tuscaloosa, AL 35487, USA}

\noindent
$^{2}${Physics Department, Columbia University, New York, NY 10027, USA}

\noindent
$^{3}${Department of Physics and Astronomy, DePauw University, Greencastle, IN 46135-0037, USA}

\noindent
$^{4}${Department of Physics and Astronomy and the Bartol Research Institute, University of Delaware, Newark, DE 19716, USA}

\noindent
$^{5}${Department of Physics and Astronomy, University of Utah, Salt Lake City, UT 84112, USA}

\noindent
$^{6}${DESY, Platanenallee 6, 15738 Zeuthen, Germany}

\noindent
$^{7}${Center for Astrophysics $|$ Harvard \& Smithsonian, Cambridge, MA 02138, USA}

\noindent
$^{8}${Physics Department, California Polytechnic State University, San Luis Obispo, CA 94307, USA}

\noindent
$^{9}${Department of Physics, Washington University, St. Louis, MO 63130, USA}

\noindent
$^{10}${School of Physics and Astronomy, University of Minnesota, Minneapolis, MN 55455, USA}

\noindent
$^{11}${Department of Physics, California State University - East Bay, Hayward, CA 94542, USA}

\noindent
$^{12}${Santa Cruz Institute for Particle Physics and Department of Physics, University of California, Santa Cruz, CA 95064, USA}

\noindent
$^{13}${Department of Physics and Astronomy, Dartmouth College, 6127 Wilder Laboratory, Hanover, NH 03755 USA}

\noindent
$^{14}${Department of Physics, University of Maryland, College Park, MD, USA }

\noindent
$^{15}${NASA GSFC, Greenbelt, MD 20771, USA}

\noindent
$^{16}${School of Natural Sciences, University of Galway, University Road, Galway, H91 TK33, Ireland}

\noindent
$^{17}${Physics Department, McGill University, Montreal, QC H3A 2T8, Canada}

\noindent
$^{18}${School of Physics, University College Dublin, Belfield, Dublin 4, Ireland}

\noindent
$^{19}${Department of Physics and Astronomy, University of Iowa, Van Allen Hall, Iowa City, IA 52242, USA}

\noindent
$^{20}${Department of Physics and Astronomy, Barnard College, Columbia University, NY 10027, USA}

\noindent
$^{21}${ Arthur B. McDonald Canadian Astroparticle Physics Research Institute, 64 Bader Lane, Queen's University, Kingston, ON Canada, K7L 3N6}

\noindent
$^{22}${Department of Physics and Astronomy, University of California, Los Angeles, CA 90095, USA}

\noindent
$^{23}${Department of Physics, Engineering Physics and Astronomy, Queen's University, Kingston, ON K7L 3N6, Canada}

\noindent
$^{24}${Institute of Physics and Astronomy, University of Potsdam, 14476 Potsdam-Golm, Germany}

\noindent
$^{25}${Department of Physical Sciences, Munster Technological University, Bishopstown, Cork, T12 P928, Ireland}

\noindent
$^{26}${Department of Physics and Astronomy, Purdue University, West Lafayette, IN 47907, USA}

\noindent
$^{27}${Department of Physics, Indiana University-Purdue University Indianapolis, Indianapolis, IN 46202, USA}

\noindent
$^{28}${Department of Physics and Astronomy, Iowa State University, Ames, IA 50011, USA}

\noindent
$^{29}${Columbia Astrophysics Laboratory, Columbia University, New York, NY 10027, USA}

\end{document}